\documentclass[proof]{WileyASNA-v1}

\articletype{Article Type}%

\received{6 September 2019}
\revised{XX September 2019}
\accepted{XX September 2019}

\begin{document}

\title{Lessons learned from 19 years of high-resolution
X-ray spectroscopy of galaxy clusters with RGS\protect\thanks{RGS is the Reflection Grating Spectrometer on board XMM-\textit{Newton}.}}

\author[1,2]{Ciro Pinto*}

\author[3]{Andrew C. Fabian}

\author[4]{Jeremy S. Sanders}

\author[5]{Jelle de Plaa}

\authormark{C. Pinto \textsc{et al}}

\address[1]{\orgdiv{ESTEC}, \orgname{ESA}, \orgaddress{Keplerlaan 1, 2201AZ Noordwijk, \country{Netherlands}}}

\address[2]{\orgdiv{IASF Palermo}, \orgname{INAF}, \orgaddress{Via U. La Malfa 153, I-90146 Palermo, \country{Italy}}}

\address[3]{\orgname{Institute of Astronomy}, \orgaddress{Cambridge, CB3 0HA\state{}, \country{United Kingdom}}}

\address[4]{\orgname{Max-Planck-Institut fur extraterrestrische Physik}, \orgaddress{Giessenbachstrasse 1, D-85748 Garching, \country{Germany}}}

\address[5]{\orgname{Netherlands Institute for Space Research}, \orgaddress{Sorbonnelaan 2, 3584 CA Utrecht, \country{Netherlands}}}

\corres{*Ciro Pinto, IASF Palermo, INAF, Via U. La Malfa 153, I-90146 Palermo, Italy. \email{ciro.pinto@inaf.it}}


\abstract{The intracluster medium (ICM) contains the vast majority of the baryonic matter in galaxy clusters and is heated to X-ray radiating temperatures. X-ray spectroscopy is therefore a key to understand both the morphology and the dynamics of galaxy clusters. Here we recall crucial evolutionary problems of galaxy clusters unveiled by 19 years of high-resolution X-ray spectroscopy with the Reflection Grating Spectrometer (RGS) on board XMM-\textit{Newton}. Its exquisite combination of effective area, spectral resolution and excellent performance over two decades enabled transformational science and important discoveries such as the lack of strong cooling flows, the constraints on ICM turbulence and cooling-heating balance. The ability of RGS to resolve individual ICM spectral lines reveals in great detail the chemical enrichment in clusters by supernovae and AGB stars. RGS spectra clearly showed that the ICM plasma is overall in thermal equilibrium which is unexpected given the wealth of energetic phenomena such as jets from supermassive black holes and mergers.}


\jnlcitation{\cname{%
\author{C. Pinto}, 
\author{J. S. Sanders}, 
\author{A. C. Fabian}, and
\author{J. de Plaa}} (\cyear{2019}), 
\ctitle{Lessons learned from 19 years of high-resolution X-ray spectroscopy of galaxy clusters with RGS}, \cjournal{Astron. Nachr.}, \cvol{2019;00:1--6}.}


\maketitle


\section{Introduction}\label{sec:intro}

The intracluster medium (ICM) embedded in the deep gravitational potentials of galaxy clusters
is a unique laboratory where highly energetic astrophysical phenomena occur.
Its thermodynamic and chemical properties witness the evolution of 
the individual galaxies as altered by several phenomena such as galaxy mergers,
gas sloshing and feedback from active galactic nuclei (AGN).
About a third of galaxy clusters shows
a highly peaked density profile,
which corresponds to the region where the central cooling time is significantly shorter
than the age of the Universe and that of the clusters 
(cool core clusters, e.g. \citealt{Hudson2010}).
In the absence of heating, this would imply the cooling of hundreds of solar masses 
of gas per year below $10^{6}$\,K \citep{Fabian1994} for the massive clusters,
with a consequent star formation rate of a similar order of magnitude.
Spectral models of massive cooling flows of 
100s\,M$_{\odot}\,{\rm yr}^{-1}$ were consistent with
the shapes of spectra from low-resolution X-ray spectrometers on board 
early observatories like \textit{ROSAT}, although some clusters
showed breaks in the mass deposition rates
(see e.g. \citealt{Peres1998} and references therein).
Absorption from cool gas within the cluster was invoked to explain some of the 
missing emission at the soft X-ray energies.

Star formation triggered by gas cooling is expected to enhance the metallicity of the 
intracluster medium and to complicate its chemical structure. 
Core-collapse supernovae are known to contribute to lighter elements such as
O, Ne and Mg, while type Ia supernovae dominate the fraction of S, Fe, Ni and other
heavy elements (see e.g. \citealt{dePlaa2007}). Nitrogen is most likely produced by
AGB stars. The relative ratios of N, O and Ne to Fe (among the most abundant metals) are
therefore the means to determine the star formation history of clusters.

Nowadays, the most commonly used X-ray spectrometers are charge-coupled device (CCD) cameras
like those on board XMM-\textit{Newton}, \textit{Chandra} and \textit{Suzaku}
owing to their high effective area although low-to-medium
spectral resolution ($R=E / \Delta E \sim 10-50$). These detectors are very useful 
due to their 2D imagining capabilities and high-count-rate spectra, which
enabled important discoveries such as the bubbles inflated by AGN jets 
(witnessing the effects of the supermassive 
black hole of the central brightest cluster galaxy, BCG, 
onto the surrounding ICM, e.g. \citealt{Churazov2000} 
and \citealt{Fabian2005}) and the flat radial profiles of the abundances 
(suggesting an early enrichment or
a complex mixing / cycle of the metals within the cluster, see e.g. \citealt{Matsushita2007a} and \citealt{Urban2017}). 

However, CCD detectors are not able to detect and resolve individual X-ray 
emission lines, particularly those produced by high-ionisation ions of nitrogen 
(mainly N\,{\small VII}), 
oxygen (O\,{\small VII-VIII}), neon (Ne\,{\small IX-X}) and the Fe L complex 
(Fe\,{\small XXIV} or lower ionisation states), which are the main tracers
of cooling flows.
This causes large uncertainties in the estimates of temperatures and emissivities
of cool gas phases and amount of cool gas. 

High energy-resolution 
dispersive spectrometers (gratings) and micro-calorimeters were designed to 
constrain cooling flows and search for
high levels of turbulent and bulk motions in the intracluster gas, which were expected given
the wealth of highly energetic phenomena occurring in the cores and in the outskirts of 
clusters. Galactic mergers, gas sloshing and AGN jets are thought to generate
motions for up to $\sim500-1000$ km s$^{-1}$ (see e.g. \citealt{Ascasibar2006}; 
\citealt{Lau2009} and \citealt{Bruggen2005}).
This might release enough heat to balance cooling.
The level of turbulence is crucial to identify any bias in  
mass measurements of clusters due to the assumption of hydrostatic equilibrium.


\section{XMM-\textit{Newton} / RGS}\label{sec:XMM_RGS}


Detailed astrophysics of X-ray sources
often requires significantly higher spectral resolution ($R=E / \Delta E \sim 100-1000$). 
This has been offered, excluding the brief life of \textit{Hitomi} (operating for five weeks
and carrying a microcalorimeter, \citealt{Hitomi2016nat})
only by the grating spectrometers on board XMM-\textit{Newton} 
(the Reflection Grating Spectrometer, RGS, see \citealt{denherder2001}) 
and \textit{Chandra} (the low / high 
transmission grating spectrometers, LETGS and HETGS). 

Gratings have less applicability than CCDs due to limited imaging (mostly 1D) and 
count-rate (lower effective area) capabilities. They are therefore optimal
for bright sources (Flux$\,_{0.3-2 \, \rm keV} \gtrsim 10^{-12} {\rm \, erg \, s^{-1} \, cm}^{-2}$) and, 
particularly, those with small angular size ($\lesssim1'$) since the gratings do not have a slit.
For instance, the RGS spectral lines are broadened according to the law
$\Delta\lambda = 0.138 \, \Delta\theta \, {\mbox{\AA}} / m,$
where $\Delta\lambda$ is the wavelength broadening, 
$\Delta\theta$ is the source extent in arc minutes
and $m$ is the spectral order.
RGS is currently the ideal grating spectrometer for extended 
sources owing to its high spectral resolution ($R=E / \Delta E \sim 100-800$) and 
sufficient effective area ($\sim20-100$ cm$^{2}$) in the soft X-ray energy range 
($\sim 0.33-1.77$ keV) where a forest of spectral lines are produced by the K shells
of some among the most abundant elements in the Universe such as C, N, O, Ne and Mg
and the complex L shells of Fe and Ni.

\section{RGS unique contributions}\label{sec:RGS_unique}

We first focus on the major results in the astrophysics of galaxy clusters
primarily driven by the capabilities of the RGS before moving to the synergies with other facilities.

\subsection{Cooling rates}\label{sec:RGS_coolrates}

\begin{figure}[t]
\centerline{\includegraphics[width=0.975\columnwidth, angle=0]{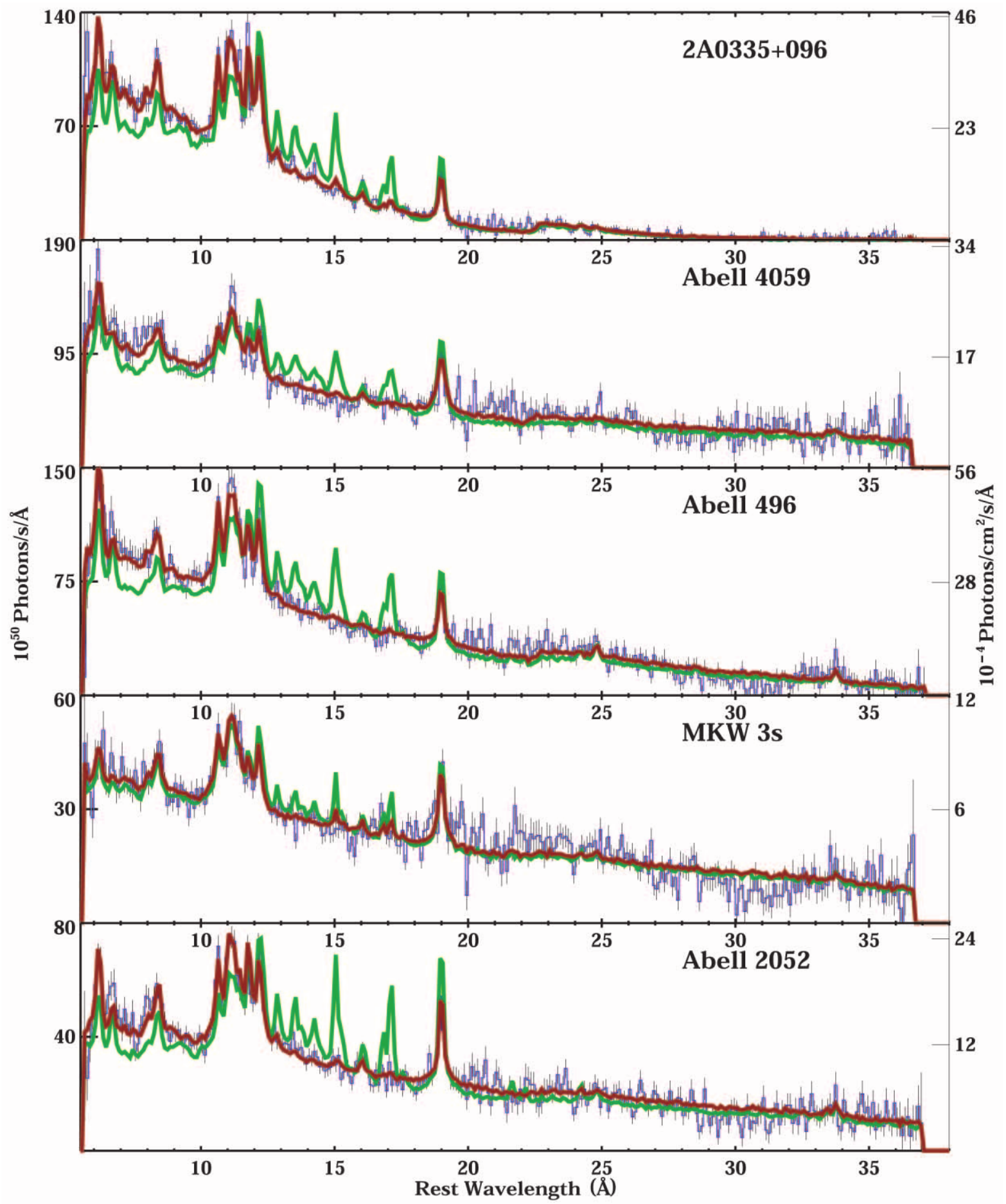}}
\vspace{-0.3cm}
\caption{First RGS observations of galaxy clusters \citep{Peterson2003}.
Comparison of the data (blue), the empirical best fit model (red), and the standard cooling-flow model (green). 
The latter overestimates emission lines from cool gas.\label{fig:Peterson2003}}
\vspace{-0.3cm}
\end{figure}

The first light of RGS onto clusters of galaxies brought up a huge surprise.
The observed cooling rates of cool-core clusters were much lower than the theoretical predictions
and the previous measurements of CCD detectors, showing only
modest levels of a few dozens M$_{\odot}\,{\rm yr}^{-1}$ (see, e.g., 
\citealt{Kaastra2001,Peterson2001,Tamura2001b}). 
Fig.\,\ref{fig:Peterson2003} shows a comparison between the RGS spectra of some clusters 
with empirical (quasi-isothermal) models of gas in collisional equilibrium and cooling-flow models.
The cooling-flow model is not a best fit, but calculated by merely taking the soft X-ray flux 
in the empirical model and using the standard isobaric temperature distribution. 
The cooling-flow models overpredict the strength of the Fe\,{\small XVII-XVIII} emission lines from 
the plasma at low temperatures. 
There is a remarkable lack of gas below $\sim1-2$ keV
(see also \citealt{Sanders2008a} and \citealt{Liu2019}). 
Cooling rates are even lower below 0.5 keV as shown by the faint
O\,{\small VII} lines recently discovered in elliptical galaxies and clusters
 (\citealt{Pinto2014,Pinto2016mnras}).
In Sect.\,\ref{sec:RGS_synergies_longwavelengths} 
we also show the comparison between the cooling rates
measured in clusters through the RGS with indicators of star formation rates determined
with facilities at low energies.

The X-ray emission lines in RGS spectra showed that the ICM plasma is overall in thermal 
equilibrium which was not obvious given the presence of energetic phenomena such as jets 
from supermassive black holes, mergers and cooling flows.

\subsection{Kinematics}\label{sec:RGS_velocities}

Only since 2010s RGS has been used to place constraints on turbulence by measuring 
the velocity dispersion of the ICM, mainly
due to the line broadening caused by the
spatial extent of clusters. 
However, the line spatial broadening of clusters
with X-ray core $\lesssim1'$ is limited to a few hundred
km s$^{-1}$ and can be corrected through CCD surface brightness profiles.

\citet{Sanders2010} placed the first 90\% upper limit (274\,km\,s$^{-1}$) on the velocity broadening
of the luminous cool-core cluster A\,1835 at redshift 0.25 with RGS spectra. 
\citet{Bulbul2012} constrained the turbulent motions in the compact core of 
A\,3112 to be lower than 206\,km\,s$^{-1}$.
\citet{Sanders2011b} found turbulent broadening below 700\,km\,s$^{-1}$ for 30 
clusters, groups, and elliptical galaxies observed with XMM-\textit{Newton}/RGS,
subsequently confirmed by
\citet{Pinto2015} on nearby ($z\lesssim0.08$) clusters using 
the CHEERS 
sample of 44 sources.
They also showed that the upper limits on the Mach numbers are larger
than the values required to balance cooling, suggesting that dissipation of turbulence 
may be high enough to heat the gas and prevent gas cooling (if turbulence is 
locally replenished).

Similar levels of turbulence have been constrained in giant elliptical galaxies with resonant scattering
(see, e.g., \citealt{Werner2009}, \citealt{dePlaa2012}, \citealt{Pinto2016mnras}, \citealt{Ogorzalek2017}).
When turbulence is low, the Fe\,{\small{XVII}} resonant line ($\lambda =$ 15\,{\AA}) is optically thick and
suppressed along the line of sight, while the 17\,{\AA} forbidden line remains optically thin. The comparison
of their observed line ratio with simulations for different Mach numbers constrains the level of turbulence.
This method is very efficient for cool ($kT<0.9$\,keV) giant elliptical galaxies rich in Fe\,{\small{XVII}}
emission lines, but it is affected by the systematic uncertainty (up to $\sim$20\%) 
in the line ratio. Currently, we cannot use this technique for clusters because they typically have higher ionisation lines (e.g. Fe\,{\small{XXV}}), which fall out of the RGS energy band. 
In a few years, studies of resonant scattering in clusters will be possible with \textit{XRISM} (\citealt{Guainazzi2018}) as its precursor, \textit{Hitomi}, did for the sole observation of the Perseus cluster \citep{Hitomi2017rs}.
 
\begin{figure}[t]
\centerline{\includegraphics[width=0.94\columnwidth, angle=0]{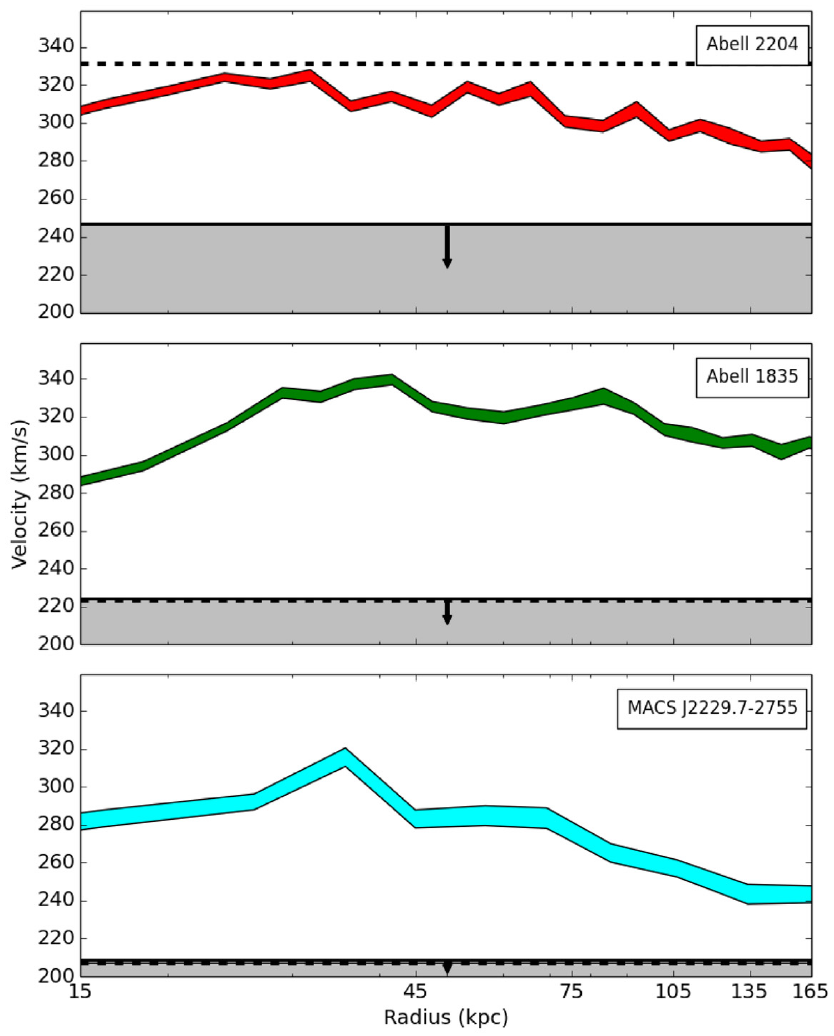}}
\vspace{-0.3cm}
\caption{Constraints on turbulent velocities for three clusters of galaxies \citep{Bambic2018}.
The coloured regions show the propagation velocity required to balance radiative cooling (1$\sigma$ error). 
The dashed (solid) line shows the 90\% upper limit on 1D turbulent velocities without (with) spatial 
broadening subtracted. This indicates that AGN turbulence alone is unable to balance radiative cooling in the inner core. \label{fig:Fig_Bambic18_fig3}}
\vspace{-0.3cm}
\end{figure}

Importantly, line broadening is less than $200-300$\,km\,s$^{-1}$
when the spurious spatial broadening is removed through the conversion of
CCD surface brightness profiles into line spatial broadening
\citep{Sanders2013,Pinto2015}. This indicates that turbulence contributes to the total energy 
for less than $\sim5$\,\% and is consistent with the measurements of \textit{Hitomi}
for Perseus \citep{Hitomi2016nat}. \citet{Bambic2018} and \citet{Pinto2018b} expanded 
this argument and showed that the propagation velocity is too low to achieve 
balance between gas cooling and heating via dissipation of turbulence (see Fig.\,\ref{fig:Fig_Bambic18_fig3}), 
previously invoked by \citealt{Zhuravleva2014} using surface brightness fluctuations.
An additional source of heating may be sound waves
(e.g. \citealt{Fabian2017sw}).

\subsection{Multiphaseness}\label{sec:RGS_multiphaseness}

\begin{figure}[t]
\centerline{\includegraphics[width=1.0\columnwidth, angle=0]{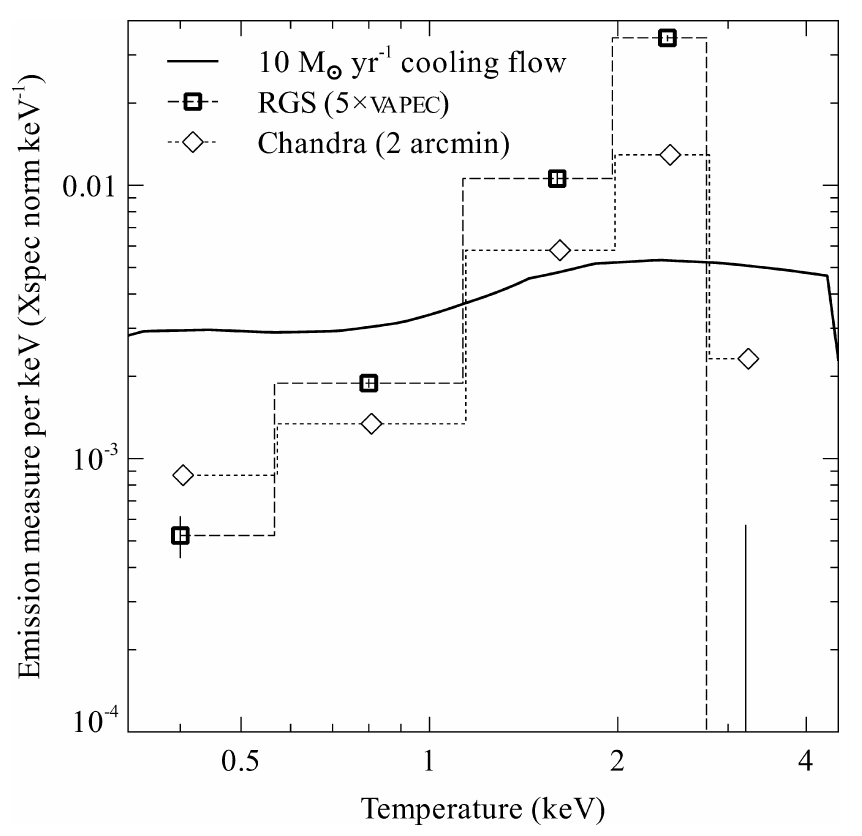}}
\vspace{-0.3cm}
\caption{Comparison of the emission measures from RGS and Chandra observations of 
 the Centaurus cluster core
\citep{Sanders2008a}. The solid line shows the expected distribution for a 
10\,M$_{\odot}$\,yr$^{-1}$ isobaric cooling flow. \label{fig:Fig_Sanders2008a_fig10}}
\vspace{-0.3cm}
\end{figure}

There was evidence for a multiphase ICM in early X-ray CCD spectra
of galaxy clusters. Spatially-resolved spectroscopy indicates a complex morphology 
and multi-temperature gas \citep{Tamura2001b, Kaastra2004, dePlaa2004, Frank2013}.
This is partly due to projection effects, because of the temperature gradient in the core. 
Deep RGS spectra of cool-core clusters provide the means from breaking some degeneracy
through the detection of individual lines that enable to distinguish between different physical models. 
The results indicate that a powerlaw
temperature distribution for the emission measure is favoured over a gaussian distribution
(see Fig.\,\ref{fig:Fig_Sanders2008a_fig10} and \citealt{Werner2006a, Sanders2008a, Liu2019}).
This is confirmed by the fact that RGS spectra can be better modelled with 2 emission components 
in collisional ionisation equilibrium rather than with a gaussian temperature
distribution (two APEC models in XSPEC\,\footnote{https://heasarc.gsfc.nasa.gov/xanadu/xspec/} or 
two CIE models in SPEX\,\footnote{https://www.sron.nl/astrophysics-spex}, 
e.g., \citealt{Pinto2015} and \citealt{dePlaa2017}) 
with the cooler component having a much lower emission measure.

\subsection{Chemical enrichment}\label{sec:RGS_chemistry}

As mentioned in Sect.\,\ref{sec:intro},
to unveil the history of chemical enrichment in clusters it is necessary to
detect and resolve individual lines from ionic species of abundant elements. 
The RGS observations showed that AGB stars dominate the enrichment of nitrogen 
due to its high abundance measured with the bright N\,{\small VII} line in the RGS spectra
of several clusters and giant elliptical galaxies
(see e.g. \citealt{Tamura2003}, \citealt{Buote2003b}, \citealt{Sanders2008a}, \citealt{Werner2006a}).
\citet{Werner2006b} and \citet{Grange2011} also measured the carbon 
abundance, confirming that the creation of nitrogen and carbon takes place 
in low- and intermediate-mass stars (see also \citealt{dePlaa2007}).

The detections of individual O\,{\small VIII} and Ne\,{\small IX-X} 
emission lines enabled accurate measurements of $\alpha/Fe$ abundance ratios in galaxy clusters.
In most cases, the O/Fe and Ne/Fe abundance ratios - as measured with previous atomic databases -
seemed to be sub-Solar\,\footnote{`Solar' refers to the proto-Solar abundances of \citet{Lodders2009}.}
(e.g. \citealt{Tamura2003,Buote2003b,dePlaa2004,Werner2006a,Grange2011,Simionescu2009a,Bulbul2012,Mernier2016a}).
On average, the relative fractions of type Ia supernovae are 
SN Ia / (SN Ia + SN cc) $\sim 25-45$\,\%,
possibly larger than the Solar environment ($\sim15-25$\,\%) and, therefore, 
suggest additional production
of heavy elements from recent SN type Ia in the BCG.
However, the uniformity of the O/Fe and Ne/Fe abundance ratios over more than an order of magnitude
in mass range (from giant ellipticals to groups and then clusters of galaxies) and the lack of spatial 
gradients and distribution with the redshift indicate that either most metals were formed around $z \sim 2$ or
that several phenomena such as sloshing and metal uplift by AGN bubbles redistributed the metals
(see e.g. \citealt{Mernier2016b}, \citealt{dePlaa2017}).

\section{RGS synergies}\label{sec:RGS_synergies}

\subsection{Synergies with X-ray CCD detectors}\label{sec:RGS_synergies_CCD}

XMM-\textit{Newton} observations of clusters 
have showed an excellent synergy between the capabilities of RGS in measuring 
the relative abundances of light $\alpha$ elements (C, N, O, Ne and Mg) with respect to iron 
and EPIC (both pn and MOS) in determining the absolute abundances (relative to hydrogen) 
 of heavier elements (Si, S, Ar, Ca, Cr, Mn, Fe and Ni) owing to its higher
sensitivity to the bremsstrahlung continuum (see Sect.\,\ref{sec:RGS_chemistry}).

\subsection{Synergies with X-ray calorimeters}\label{sec:RGS_synergies_xraycal}

\begin{figure}[t]
\centerline{\includegraphics[width=1.0\columnwidth, angle=0]{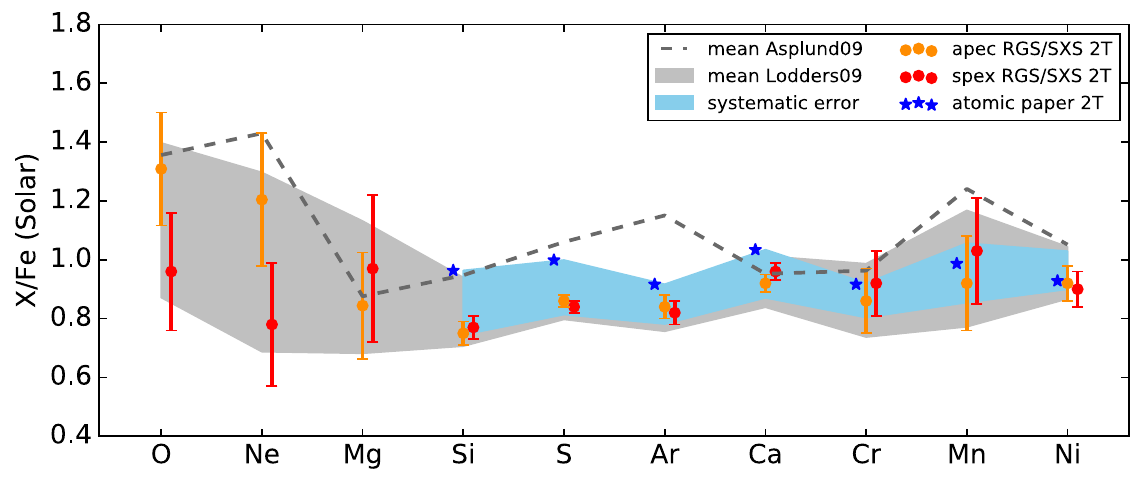}}
\caption{Abundance ratios measured for the Perseus cluster 
with XMM-\textit{Newton}/RGS (O, Ne, Mg to Fe) 
and \textit{Hitomi}/SXS (Si through Ni to Fe, \citealt{Simionescu2019a}). The light blue strip 
is the systematic uncertainty on the \textit{Hitomi} data. 
The grey strip includes both systematic and statistical uncertainties. The measured values are given with respect to the units of \citet{Lodders2009} and \citet{Asplund2009}.\label{fig:Fig_Simionescu19_fig4}}
\end{figure}

RGS is even more efficient if used together with other high resolution spectrometers
that cover the high-energy (2--10\,keV) X-ray energy band such X-ray calorimeters.
\citet{Simionescu2019a} show a remarkable example of combining the XMM-\textit{Newton}/RGS 
soft X-ray and \textit{Hitomi}/SXS hard X-ray spectra of the Perseus cluster with state-of-art atomic data,
achieving an unprecedented accuracy on the relative $\alpha$/Fe abundance ratios
(see Fig.\,\ref{fig:Fig_Simionescu19_fig4}).
The abundance pattern agrees with the Solar nebula and challenges any linear combinations 
of supernova nucleosynthesis calculations. Including neutrino physics in the yield calculations of SN cc
may improve the agreement with the observed pattern of $\alpha$ 
elements in the Perseus Cluster core.

A major improvement in the measurements of both turbulence and chemical 
abundances will be achieved with ATHENA, the most powerful X-ray mission planned
for early 2030s (see e.g. \citealt{Roncarelli2018} and \citealt{Cucchetti2018}).

\subsection{Synergies with long wavelength facilities}\label{sec:RGS_synergies_longwavelengths}

The comparison of the cooling rates measured by RGS with the results
obtained at lower energies is a key to unveil the evolution of the ICM.
For instance, \citet{Liu2019} used the RGS observations for a sub-sample of
the CHEERS catalog plus A\,1835 and
confirmed that the cooling rates are an order of magnitude lower than the 
theoretical predictions. However, they also
showed that the cooling rates may still be high enough to explain
the H$\alpha$ luminosity and the star formation rates measured with
\textit{Hubble Space Telescope} and WISE.
\citet{Pinto2018b} have shown that the cooling rate of $350\pm130\,M_{\odot}\,{\rm yr}^{-1}$ 
below 2\,keV measured with RGS in the Phoenix cluster
is consistent with the star formation rate in this object
and is high enough to produce the molecular gas found with ALMA in the filaments 
via instabilities during the buoyant rising time (see \citealt{Russell2017}).



\subsection{Improving atomic databases}\label{sec:RGS_atomic}

RGS has been crucial to test the accuracy of
atomic databases for plasmas at $\sim$ 0.1-1 keV temperatures. \citet{dePlaa2017} 
and \citet{Gu2019a} have shown that RGS spectra are sensitive enough to distinguish 
among different calculations of atomic cross sections. 
The ICM abundance pattern seems to agree with the Sun if state-of-art 
atomic databases are adopted (see Sect.\,\ref{sec:RGS_synergies_xraycal} and \citealt{Gu2019a}).
Currently, there are uncertainties for 10--20\% in the emissivities of the strongest 
emission lines. The spectral fits would appear identical in CCD spectra, but with abundances 
wrong by up to 20\%. 
Finally, RGS spectra have also been able to find the first evidence for charge exchange 
between the cold neutral gas and the hot atmosphere in galaxy clusters 
(see e.g. \citealt{Pinto2016mnras} and \citealt{Gu2018a}).

\section{Conclusions}\label{Conclusions}

\begin{figure}[t]
\centerline{\includegraphics[width=1.0\columnwidth, angle=0]{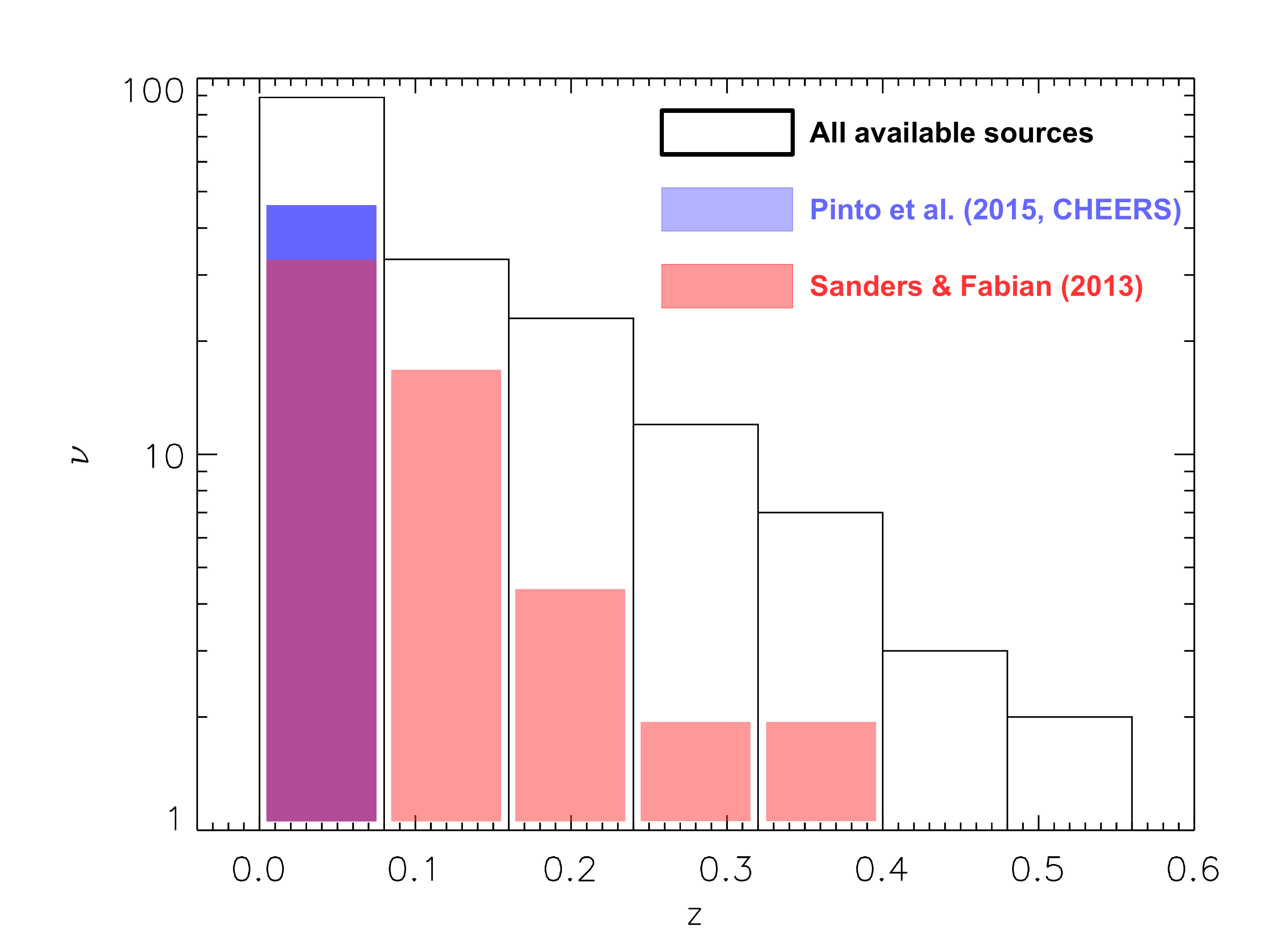}}
\vspace{-0.3cm}
\caption{Histogram of all clusters, groups and elliptical galaxies observed with RGS
at different redshift (on-axis, $t_{\rm clean} > 10$\,ks, showing Fe L and / or O\,{\small{VIII}} emission lines).\label{fig:RGScatalog}}
\vspace{-0.3cm}
\end{figure}

In almost 20 years, XMM-\textit{Newton}/RGS 
has delivered and continues to deliver fascinating and unique science. 
Here we have summarised the most crucial contribution of RGS
to the astrophysics of galaxy clusters such as the accurate measurements of cooling rates,
the constraints on turbulence and on cooling--heating balance, the accurate abundance
measurements and the supernova yields, the tests and improvements on new atomic databases.
Even now, after two decades, new scientific problems appear, triggered by the investigation of the 
available RGS spectra and the large databases. Whilst waiting for new missions, 
RGS can make significant progress by going deeper with longer exposure times
and at higher redshifts closer to the peak of the star formation in the coming decade.
Moreover there is a wealth of RGS data still to be studied (see Fig. \ref{fig:RGScatalog}). 

\section*{Acknowledgments}

This work is based on observations obtained with XMM-\textit{Newton}, an
ESA science mission funded by ESA Member States and USA (NASA).
We acknowledge support by European Space Agency (ESA) Research Fellowships.


\bibliography{Wiley-ASNA}%

\begin{thebibliography}{}

\bibitem [\protect \citeauthoryear {%
{Ascasibar}%
\ \BBA {} {Markevitch}%
}{%
{Ascasibar}%
\ \BBA {} {Markevitch}%
}{%
{\protect \APACyear {2006}}%
}]{%
Ascasibar2006}
\APACinsertmetastar {%
Ascasibar2006}%
\begin{APACrefauthors}%
{Ascasibar}, Y.%
\BCBT {}\ \BBA {} {Markevitch}, M.%
\end{APACrefauthors}%
\unskip\
\newblock
\APACrefYearMonthDay{2006}{{\APACmonth{10}}}{},
\newblock
\unskip
\newblock
\APACjournalVolNumPages{ApJ}{650}{}{102-127}.
\newblock

\PrintBackRefs{\CurrentBib}

\bibitem [\protect \citeauthoryear {%
{Asplund}%
, {Grevesse}%
, {Sauval}%
\BCBL {}\ \BBA {} {Scott}%
}{%
{Asplund}%
\ \protect \BOthers {.}}{%
{\protect \APACyear {2009}}%
}]{%
Asplund2009}
\APACinsertmetastar {%
Asplund2009}%
\begin{APACrefauthors}%
{Asplund}, M.%
, {Grevesse}, N.%
, {Sauval}, A\BPBI J.%
\BCBL {}\ \BBA {} {Scott}, P.%
\end{APACrefauthors}%
\unskip\
\newblock
\APACrefYearMonthDay{2009}{Sep}{},
\newblock
\unskip
\newblock
\APACjournalVolNumPages{\araa}{47}{1}{481-522}.
\newblock

\PrintBackRefs{\CurrentBib}

\bibitem [\protect \citeauthoryear {%
{Bambic}%
, {Pinto}%
, {Fabian}%
, {Sanders}%
\BCBL {}\ \BBA {} {Reynolds}%
}{%
{Bambic}%
\ \protect \BOthers {.}}{%
{\protect \APACyear {2018}}%
}]{%
Bambic2018}
\APACinsertmetastar {%
Bambic2018}%
\begin{APACrefauthors}%
{Bambic}, C\BPBI J.%
, {Pinto}, C.%
, {Fabian}, A\BPBI C.%
, {Sanders}, J.%
\BCBL {}\ \BBA {} {Reynolds}, C\BPBI S.%
\end{APACrefauthors}%
\unskip\
\newblock
\APACrefYearMonthDay{2018}{Jul}{},
\newblock
\unskip
\newblock
\APACjournalVolNumPages{\mnras}{478}{1}{L44-L48}.
\newblock

\PrintBackRefs{\CurrentBib}

\bibitem [\protect \citeauthoryear {%
{Br{\"u}ggen}%
, {Hoeft}%
\BCBL {}\ \BBA {} {Ruszkowski}%
}{%
{Br{\"u}ggen}%
\ \protect \BOthers {.}}{%
{\protect \APACyear {2005}}%
}]{%
Bruggen2005}
\APACinsertmetastar {%
Bruggen2005}%
\begin{APACrefauthors}%
{Br{\"u}ggen}, M.%
, {Hoeft}, M.%
\BCBL {}\ \BBA {} {Ruszkowski}, M.%
\end{APACrefauthors}%
\unskip\
\newblock
\APACrefYearMonthDay{2005}{{\APACmonth{07}}}{},
\newblock
\unskip
\newblock
\APACjournalVolNumPages{ApJ}{628}{}{153-159}.
\newblock

\PrintBackRefs{\CurrentBib}

\bibitem [\protect \citeauthoryear {%
{Bulbul}%
\ \protect \BOthers {.}}{%
{Bulbul}%
\ \protect \BOthers {.}}{%
{\protect \APACyear {2012}}%
}]{%
Bulbul2012}
\APACinsertmetastar {%
Bulbul2012}%
\begin{APACrefauthors}%
{Bulbul}, G\BPBI E.%
, {Smith}, R\BPBI K.%
, {Foster}, A.%
, {Cottam}, J.%
, {Loewenstein}, M.%
, {Mushotzky}, R.%
\BCBL {}\ \BBA {} {Shafer}, R.%
\end{APACrefauthors}%
\unskip\
\newblock
\APACrefYearMonthDay{2012}{{\APACmonth{03}}}{},
\newblock
\unskip
\newblock
\APACjournalVolNumPages{ApJ}{747}{}{32}.
\newblock

\PrintBackRefs{\CurrentBib}

\bibitem [\protect \citeauthoryear {%
{Buote}%
, {Lewis}%
, {Brighenti}%
\BCBL {}\ \BBA {} {Mathews}%
}{%
{Buote}%
\ \protect \BOthers {.}}{%
{\protect \APACyear {2003}}%
}]{%
Buote2003b}
\APACinsertmetastar {%
Buote2003b}%
\begin{APACrefauthors}%
{Buote}, D\BPBI A.%
, {Lewis}, A\BPBI D.%
, {Brighenti}, F.%
\BCBL {}\ \BBA {} {Mathews}, W\BPBI G.%
\end{APACrefauthors}%
\unskip\
\newblock
\APACrefYearMonthDay{2003}{Sep}{},
\newblock
\unskip
\newblock
\APACjournalVolNumPages{\apj}{595}{1}{151-166}.
\newblock

\PrintBackRefs{\CurrentBib}

\bibitem [\protect \citeauthoryear {%
{Churazov}%
, {Forman}%
, {Jones}%
\BCBL {}\ \BBA {} {B{\"o}hringer}%
}{%
{Churazov}%
\ \protect \BOthers {.}}{%
{\protect \APACyear {2000}}%
}]{%
Churazov2000}
\APACinsertmetastar {%
Churazov2000}%
\begin{APACrefauthors}%
{Churazov}, E.%
, {Forman}, W.%
, {Jones}, C.%
\BCBL {}\ \BBA {} {B{\"o}hringer}, H.%
\end{APACrefauthors}%
\unskip\
\newblock
\APACrefYearMonthDay{2000}{{\APACmonth{04}}}{},
\newblock
\unskip
\newblock
\APACjournalVolNumPages{A\&A}{356}{}{788-794}.
\PrintBackRefs{\CurrentBib}

\bibitem [\protect \citeauthoryear {%
{Cucchetti}%
\ \protect \BOthers {.}}{%
{Cucchetti}%
\ \protect \BOthers {.}}{%
{\protect \APACyear {2018}}%
}]{%
Cucchetti2018}
\APACinsertmetastar {%
Cucchetti2018}%
\begin{APACrefauthors}%
{Cucchetti}, E.%
, {Pointecouteau}, E.%
, {Peille}, P.%
\ et al.\end{APACrefauthors}%
\unskip\
\newblock
\APACrefYearMonthDay{2018}{Dec}{},
\newblock
\unskip
\newblock
\APACjournalVolNumPages{\aap}{620}{}{A173}.
\newblock

\PrintBackRefs{\CurrentBib}

\bibitem [\protect \citeauthoryear {%
{de Plaa}%
\ \protect \BOthers {.}}{%
{de Plaa}%
\ \protect \BOthers {.}}{%
{\protect \APACyear {2004}}%
}]{%
dePlaa2004}
\APACinsertmetastar {%
dePlaa2004}%
\begin{APACrefauthors}%
{de Plaa}, J.%
, {Kaastra}, J\BPBI S.%
, {Tamura}, T.%
, {Pointecouteau}, E.%
, {Mendez}, M.%
\BCBL {}\ \BBA {} {Peterson}, J\BPBI R.%
\end{APACrefauthors}%
\unskip\
\newblock
\APACrefYearMonthDay{2004}{Aug}{},
\newblock
\unskip
\newblock
\APACjournalVolNumPages{\aap}{423}{}{49-56}.
\newblock

\PrintBackRefs{\CurrentBib}

\bibitem [\protect \citeauthoryear {%
{de Plaa}%
\ \protect \BOthers {.}}{%
{de Plaa}%
\ \protect \BOthers {.}}{%
{\protect \APACyear {2017}}%
}]{%
dePlaa2017}
\APACinsertmetastar {%
dePlaa2017}%
\begin{APACrefauthors}%
{de Plaa}, J.%
, {Kaastra}, J\BPBI S.%
, {Werner}, N.%
\ et al.\end{APACrefauthors}%
\unskip\
\newblock
\APACrefYearMonthDay{2017}{Nov}{},
\newblock
\unskip
\newblock
\APACjournalVolNumPages{\aap}{607}{}{A98}.
\newblock

\PrintBackRefs{\CurrentBib}

\bibitem [\protect \citeauthoryear {%
{de Plaa}%
\ \protect \BOthers {.}}{%
{de Plaa}%
\ \protect \BOthers {.}}{%
{\protect \APACyear {2007}}%
}]{%
dePlaa2007}
\APACinsertmetastar {%
dePlaa2007}%
\begin{APACrefauthors}%
{de Plaa}, J.%
, {Werner}, N.%
, {Bleeker}, J\BPBI A\BPBI M.%
, {Vink}, J.%
, {Kaastra}, J\BPBI S.%
\BCBL {}\ \BBA {} {M{\'e}ndez}, M.%
\end{APACrefauthors}%
\unskip\
\newblock
\APACrefYearMonthDay{2007}{Apr}{},
\newblock
\unskip
\newblock
\APACjournalVolNumPages{\aap}{465}{2}{345-355}.
\newblock

\PrintBackRefs{\CurrentBib}

\bibitem [\protect \citeauthoryear {%
{de Plaa}%
, {Zhuravleva}%
, {Werner}%
, {Kaastra}%
\BCBL {}\ \BBA {} {Churazov}%
}{%
{de Plaa}%
\ \protect \BOthers {.}}{%
{\protect \APACyear {2012}}%
}]{%
dePlaa2012}
\APACinsertmetastar {%
dePlaa2012}%
\begin{APACrefauthors}%
{de Plaa}, J.%
, {Zhuravleva}, I.%
, {Werner}, N.%
, {Kaastra}, J\BPBI S.%
\BCBL {}\ \BBA {} {Churazov}, E\BPBI e\BPBI a.%
\end{APACrefauthors}%
\unskip\
\newblock
\APACrefYearMonthDay{2012}{{\APACmonth{03}}}{},
\newblock
\unskip
\newblock
\APACjournalVolNumPages{A\&A}{539}{}{A34}.
\newblock

\PrintBackRefs{\CurrentBib}

\bibitem [\protect \citeauthoryear {%
{den Herder}%
\ \protect \BOthers {.}}{%
{den Herder}%
\ \protect \BOthers {.}}{%
{\protect \APACyear {2001}}%
}]{%
denherder2001}
\APACinsertmetastar {%
denherder2001}%
\begin{APACrefauthors}%
{den Herder}, J\BPBI W.%
, {Brinkman}, A\BPBI C.%
, {Kahn}, S\BPBI M.%
\ et al.\end{APACrefauthors}%
\unskip\
\newblock
\APACrefYearMonthDay{2001}{Jan}{},
\newblock
\unskip
\newblock
\APACjournalVolNumPages{\aap}{365}{}{L7-L17}.
\newblock

\PrintBackRefs{\CurrentBib}

\bibitem [\protect \citeauthoryear {%
{Fabian}%
}{%
{Fabian}%
}{%
{\protect \APACyear {1994}}%
}]{%
Fabian1994}
\APACinsertmetastar {%
Fabian1994}%
\begin{APACrefauthors}%
{Fabian}, A\BPBI C.%
\end{APACrefauthors}%
\unskip\
\newblock
\APACrefYearMonthDay{1994}{}{},
\newblock
\unskip
\newblock
\APACjournalVolNumPages{ARA\&A}{32}{}{277-318}.
\newblock

\PrintBackRefs{\CurrentBib}

\bibitem [\protect \citeauthoryear {%
{Fabian}%
, {Sanders}%
, {Taylor}%
\BCBL {}\ \BBA {} {Allen}%
}{%
{Fabian}%
\ \protect \BOthers {.}}{%
{\protect \APACyear {2005}}%
}]{%
Fabian2005}
\APACinsertmetastar {%
Fabian2005}%
\begin{APACrefauthors}%
{Fabian}, A\BPBI C.%
, {Sanders}, J\BPBI S.%
, {Taylor}, G\BPBI B.%
\BCBL {}\ \BBA {} {Allen}, S\BPBI W.%
\end{APACrefauthors}%
\unskip\
\newblock
\APACrefYearMonthDay{2005}{{\APACmonth{06}}}{},
\newblock
\unskip
\newblock
\APACjournalVolNumPages{MNRAS}{360}{}{L20-L24}.
\newblock

\PrintBackRefs{\CurrentBib}

\bibitem [\protect \citeauthoryear {%
{Fabian}%
\ \protect \BOthers {.}}{%
{Fabian}%
\ \protect \BOthers {.}}{%
{\protect \APACyear {2017}}%
}]{%
Fabian2017sw}
\APACinsertmetastar {%
Fabian2017sw}%
\begin{APACrefauthors}%
{Fabian}, A\BPBI C.%
, {Walker}, S\BPBI A.%
, {Russell}, H\BPBI R.%
, {Pinto}, C.%
, {Sanders}, J\BPBI S.%
\BCBL {}\ \BBA {} {Reynolds}, C\BPBI S.%
\end{APACrefauthors}%
\unskip\
\newblock
\APACrefYearMonthDay{2017}{{\APACmonth{01}}}{},
\newblock
\unskip
\newblock
\APACjournalVolNumPages{MNRAS}{464}{}{L1-L5}.
\newblock

\PrintBackRefs{\CurrentBib}

\bibitem [\protect \citeauthoryear {%
{Frank}%
, {Peterson}%
, {Andersson}%
, {Fabian}%
\BCBL {}\ \BBA {} {Sanders}%
}{%
{Frank}%
\ \protect \BOthers {.}}{%
{\protect \APACyear {2013}}%
}]{%
Frank2013}
\APACinsertmetastar {%
Frank2013}%
\begin{APACrefauthors}%
{Frank}, K\BPBI A.%
, {Peterson}, J\BPBI R.%
, {Andersson}, K.%
, {Fabian}, A\BPBI C.%
\BCBL {}\ \BBA {} {Sanders}, J\BPBI S.%
\end{APACrefauthors}%
\unskip\
\newblock
\APACrefYearMonthDay{2013}{{\APACmonth{02}}}{},
\newblock
\unskip
\newblock
\APACjournalVolNumPages{ApJ}{764}{}{46}.
\newblock

\PrintBackRefs{\CurrentBib}

\bibitem [\protect \citeauthoryear {%
{Grange}%
\ \protect \BOthers {.}}{%
{Grange}%
\ \protect \BOthers {.}}{%
{\protect \APACyear {2011}}%
}]{%
Grange2011}
\APACinsertmetastar {%
Grange2011}%
\begin{APACrefauthors}%
{Grange}, Y\BPBI G.%
, {de Plaa}, J.%
, {Kaastra}, J\BPBI S.%
, {Werner}, N.%
, {Verbunt}, F.%
, {Paerels}, F.%
\BCBL {}\ \BBA {} {de Vries}, C\BPBI P.%
\end{APACrefauthors}%
\unskip\
\newblock
\APACrefYearMonthDay{2011}{Jul}{},
\newblock
\unskip
\newblock
\APACjournalVolNumPages{\aap}{531}{}{A15}.
\newblock

\PrintBackRefs{\CurrentBib}

\bibitem [\protect \citeauthoryear {%
{Gu}%
\ \protect \BOthers {.}}{%
{Gu}%
\ \protect \BOthers {.}}{%
{\protect \APACyear {2018}}%
}]{%
Gu2018a}
\APACinsertmetastar {%
Gu2018a}%
\begin{APACrefauthors}%
{Gu}, L.%
, {Mao}, J.%
, {de Plaa}, J.%
, {Raassen}, A\BPBI J\BPBI J.%
, {Shah}, C.%
\BCBL {}\ \BBA {} {Kaastra}, J\BPBI S.%
\end{APACrefauthors}%
\unskip\
\newblock
\APACrefYearMonthDay{2018}{Mar}{},
\newblock
\unskip
\newblock
\APACjournalVolNumPages{\aap}{611}{}{A26}.
\newblock

\PrintBackRefs{\CurrentBib}

\bibitem [\protect \citeauthoryear {%
{Gu}%
\ \protect \BOthers {.}}{%
{Gu}%
\ \protect \BOthers {.}}{%
{\protect \APACyear {2019}}%
}]{%
Gu2019a}
\APACinsertmetastar {%
Gu2019a}%
\begin{APACrefauthors}%
{Gu}, L.%
, {Raassen}, A\BPBI J\BPBI J.%
, {Mao}, J.%
\ et al.\end{APACrefauthors}%
\unskip\
\newblock
\APACrefYearMonthDay{2019}{Jul}{},
\newblock
\unskip
\newblock
\APACjournalVolNumPages{\aap}{627}{}{A51}.
\newblock

\PrintBackRefs{\CurrentBib}

\bibitem [\protect \citeauthoryear {%
{Guainazzi}%
\ \BBA {} {Tashiro}%
}{%
{Guainazzi}%
\ \BBA {} {Tashiro}%
}{%
{\protect \APACyear {2018}}%
}]{%
Guainazzi2018}
\APACinsertmetastar {%
Guainazzi2018}%
\begin{APACrefauthors}%
{Guainazzi}, M.%
\BCBT {}\ \BBA {} {Tashiro}, M\BPBI S.%
\end{APACrefauthors}%
\unskip\
\newblock
\APACrefYearMonthDay{2018}{{\APACmonth{07}}}{},
\newblock
\unskip
\newblock
\APACjournalVolNumPages{ArXiv e-prints}{}{}{}.
\PrintBackRefs{\CurrentBib}

\bibitem [\protect \citeauthoryear {%
{Hitomi Collaboration}%
}{%
{Hitomi Collaboration}%
}{%
{\protect \APACyear {2016}}%
}]{%
Hitomi2016nat}
\APACinsertmetastar {%
Hitomi2016nat}%
\begin{APACrefauthors}%
{Hitomi Collaboration}.%
\end{APACrefauthors}%
\unskip\
\newblock
\APACrefYearMonthDay{2016}{{\APACmonth{07}}}{},
\newblock
\unskip
\newblock
\APACjournalVolNumPages{Nature}{535}{}{117-121}.
\newblock

\PrintBackRefs{\CurrentBib}

\bibitem [\protect \citeauthoryear {%
{Hitomi Collaboration}%
}{%
{Hitomi Collaboration}%
}{%
{\protect \APACyear {2017}}%
}]{%
Hitomi2017rs}
\APACinsertmetastar {%
Hitomi2017rs}%
\begin{APACrefauthors}%
{Hitomi Collaboration}.%
\end{APACrefauthors}%
\unskip\
\newblock
\APACrefYearMonthDay{2017}{{\APACmonth{10}}}{},
\newblock
\unskip
\newblock
\APACjournalVolNumPages{ArXiv e-prints}{}{}{}.
\PrintBackRefs{\CurrentBib}

\bibitem [\protect \citeauthoryear {%
{Hudson}%
\ \protect \BOthers {.}}{%
{Hudson}%
\ \protect \BOthers {.}}{%
{\protect \APACyear {2010}}%
}]{%
Hudson2010}
\APACinsertmetastar {%
Hudson2010}%
\begin{APACrefauthors}%
{Hudson}, D\BPBI S.%
, {Mittal}, R.%
, {Reiprich}, T\BPBI H.%
, {Nulsen}, P\BPBI E\BPBI J.%
, {Andernach}, H.%
\BCBL {}\ \BBA {} {Sarazin}, C\BPBI L.%
\end{APACrefauthors}%
\unskip\
\newblock
\APACrefYearMonthDay{2010}{{\APACmonth{04}}}{},
\newblock
\unskip
\newblock
\APACjournalVolNumPages{ApJ}{513}{}{A37}.
\newblock

\PrintBackRefs{\CurrentBib}

\bibitem [\protect \citeauthoryear {%
{Kaastra}%
\ \protect \BOthers {.}}{%
{Kaastra}%
\ \protect \BOthers {.}}{%
{\protect \APACyear {2001}}%
}]{%
Kaastra2001}
\APACinsertmetastar {%
Kaastra2001}%
\begin{APACrefauthors}%
{Kaastra}, J\BPBI S.%
, {Ferrigno}, C.%
, {Tamura}, T.%
, {Paerels}, F\BPBI B\BPBI S.%
, {Peterson}, J\BPBI R.%
\BCBL {}\ \BBA {} {Mittaz}, J\BPBI P\BPBI D.%
\end{APACrefauthors}%
\unskip\
\newblock
\APACrefYearMonthDay{2001}{Jan}{},
\newblock
\unskip
\newblock
\APACjournalVolNumPages{\aap}{365}{}{L99-L103}.
\newblock

\PrintBackRefs{\CurrentBib}

\bibitem [\protect \citeauthoryear {%
{Kaastra}%
\ \protect \BOthers {.}}{%
{Kaastra}%
\ \protect \BOthers {.}}{%
{\protect \APACyear {2004}}%
}]{%
Kaastra2004}
\APACinsertmetastar {%
Kaastra2004}%
\begin{APACrefauthors}%
{Kaastra}, J\BPBI S.%
, {Tamura}, T.%
, {Peterson}, J\BPBI R.%
\ et al.\end{APACrefauthors}%
\unskip\
\newblock
\APACrefYearMonthDay{2004}{Jan}{},
\newblock
\unskip
\newblock
\APACjournalVolNumPages{\aap}{413}{}{415-439}.
\newblock

\PrintBackRefs{\CurrentBib}

\bibitem [\protect \citeauthoryear {%
{Lau}%
, {Kravtsov}%
\BCBL {}\ \BBA {} {Nagai}%
}{%
{Lau}%
\ \protect \BOthers {.}}{%
{\protect \APACyear {2009}}%
}]{%
Lau2009}
\APACinsertmetastar {%
Lau2009}%
\begin{APACrefauthors}%
{Lau}, E\BPBI T.%
, {Kravtsov}, A\BPBI V.%
\BCBL {}\ \BBA {} {Nagai}, D.%
\end{APACrefauthors}%
\unskip\
\newblock
\APACrefYearMonthDay{2009}{{\APACmonth{11}}}{},
\newblock
\unskip
\newblock
\APACjournalVolNumPages{ApJ}{705}{}{1129-1138}.
\newblock

\PrintBackRefs{\CurrentBib}

\bibitem [\protect \citeauthoryear {%
{Liu}%
, {Pinto}%
, {Fabian}%
, {Russell}%
\BCBL {}\ \BBA {} {Sanders}%
}{%
{Liu}%
\ \protect \BOthers {.}}{%
{\protect \APACyear {2019}}%
}]{%
Liu2019}
\APACinsertmetastar {%
Liu2019}%
\begin{APACrefauthors}%
{Liu}, H.%
, {Pinto}, C.%
, {Fabian}, A\BPBI C.%
, {Russell}, H\BPBI R.%
\BCBL {}\ \BBA {} {Sanders}, J\BPBI S.%
\end{APACrefauthors}%
\unskip\
\newblock
\APACrefYearMonthDay{2019}{May}{},
\newblock
\unskip
\newblock
\APACjournalVolNumPages{\mnras}{485}{2}{1757-1774}.
\newblock

\PrintBackRefs{\CurrentBib}

\bibitem [\protect \citeauthoryear {%
{Lodders}%
\ \BBA {} {Palme}%
}{%
{Lodders}%
\ \BBA {} {Palme}%
}{%
{\protect \APACyear {2009}}%
}]{%
Lodders2009}
\APACinsertmetastar {%
Lodders2009}%
\begin{APACrefauthors}%
{Lodders}, K.%
\BCBT {}\ \BBA {} {Palme}, H.%
\end{APACrefauthors}%
\unskip\
\newblock
\APACrefYearMonthDay{2009}{{\APACmonth{09}}}{},
\newblock
\unskip
\newblock
\APACjournalVolNumPages{Meteoritics and Planetary Science
  Supplement}{72}{}{5154-+}.
\PrintBackRefs{\CurrentBib}

\bibitem [\protect \citeauthoryear {%
{Matsushita}%
\ \protect \BOthers {.}}{%
{Matsushita}%
\ \protect \BOthers {.}}{%
{\protect \APACyear {2007}}%
}]{%
Matsushita2007a}
\APACinsertmetastar {%
Matsushita2007a}%
\begin{APACrefauthors}%
{Matsushita}, K.%
, {Fukazawa}, Y.%
, {Hughes}, J\BPBI P.%
\ et al.\end{APACrefauthors}%
\unskip\
\newblock
\APACrefYearMonthDay{2007}{Jan}{},
\newblock
\unskip
\newblock
\APACjournalVolNumPages{\pasj}{59}{}{327-338}.
\newblock

\PrintBackRefs{\CurrentBib}

\bibitem [\protect \citeauthoryear {%
{Mernier}%
\ \protect \BOthers {.}}{%
{Mernier}%
\ \protect \BOthers {.}}{%
{\protect \APACyear {2016}}%
{\protect \APACexlab {{\protect \BCnt {1}}}}}]{%
Mernier2016b}
\APACinsertmetastar {%
Mernier2016b}%
\begin{APACrefauthors}%
{Mernier}, F.%
, {de Plaa}, J.%
, {Pinto}, C.%
\ et al.\end{APACrefauthors}%
\unskip\
\newblock
\APACrefYearMonthDay{2016{\protect \BCnt {1}}}{Nov}{},
\newblock
\unskip
\newblock
\APACjournalVolNumPages{\aap}{595}{}{A126}.
\newblock

\PrintBackRefs{\CurrentBib}

\bibitem [\protect \citeauthoryear {%
{Mernier}%
\ \protect \BOthers {.}}{%
{Mernier}%
\ \protect \BOthers {.}}{%
{\protect \APACyear {2016}}%
{\protect \APACexlab {{\protect \BCnt {2}}}}}]{%
Mernier2016a}
\APACinsertmetastar {%
Mernier2016a}%
\begin{APACrefauthors}%
{Mernier}, F.%
, {de Plaa}, J.%
, {Pinto}, C.%
\ et al.\end{APACrefauthors}%
\unskip\
\newblock
\APACrefYearMonthDay{2016{\protect \BCnt {2}}}{Aug}{},
\newblock
\unskip
\newblock
\APACjournalVolNumPages{\aap}{592}{}{A157}.
\newblock

\PrintBackRefs{\CurrentBib}

\bibitem [\protect \citeauthoryear {%
{Ogorzalek}%
\ \protect \BOthers {.}}{%
{Ogorzalek}%
\ \protect \BOthers {.}}{%
{\protect \APACyear {2017}}%
}]{%
Ogorzalek2017}
\APACinsertmetastar {%
Ogorzalek2017}%
\begin{APACrefauthors}%
{Ogorzalek}, A.%
, {Zhuravleva}, I.%
, {Allen}, S\BPBI W.%
\ et al.\end{APACrefauthors}%
\unskip\
\newblock
\APACrefYearMonthDay{2017}{{\APACmonth{12}}}{},
\newblock
\unskip
\newblock
\APACjournalVolNumPages{MNRAS}{472}{}{1659-1676}.
\newblock

\PrintBackRefs{\CurrentBib}

\bibitem [\protect \citeauthoryear {%
{Peres}%
\ \protect \BOthers {.}}{%
{Peres}%
\ \protect \BOthers {.}}{%
{\protect \APACyear {1998}}%
}]{%
Peres1998}
\APACinsertmetastar {%
Peres1998}%
\begin{APACrefauthors}%
{Peres}, C\BPBI B.%
, {Fabian}, A\BPBI C.%
, {Edge}, A\BPBI C.%
, {Allen}, S\BPBI W.%
, {Johnstone}, R\BPBI M.%
\BCBL {}\ \BBA {} {White}, D\BPBI A.%
\end{APACrefauthors}%
\unskip\
\newblock
\APACrefYearMonthDay{1998}{Aug}{},
\newblock
\unskip
\newblock
\APACjournalVolNumPages{\mnras}{298}{2}{416-432}.
\newblock

\PrintBackRefs{\CurrentBib}

\bibitem [\protect \citeauthoryear {%
{Peterson}%
\ \protect \BOthers {.}}{%
{Peterson}%
\ \protect \BOthers {.}}{%
{\protect \APACyear {2003}}%
}]{%
Peterson2003}
\APACinsertmetastar {%
Peterson2003}%
\begin{APACrefauthors}%
{Peterson}, J\BPBI R.%
, {Kahn}, S\BPBI M.%
, {Paerels}, F\BPBI B\BPBI S.%
\ et al.\end{APACrefauthors}%
\unskip\
\newblock
\APACrefYearMonthDay{2003}{{\APACmonth{06}}}{},
\newblock
\unskip
\newblock
\APACjournalVolNumPages{A\&A}{590}{}{207-224}.
\newblock

\PrintBackRefs{\CurrentBib}

\bibitem [\protect \citeauthoryear {%
{Peterson}%
\ \protect \BOthers {.}}{%
{Peterson}%
\ \protect \BOthers {.}}{%
{\protect \APACyear {2001}}%
}]{%
Peterson2001}
\APACinsertmetastar {%
Peterson2001}%
\begin{APACrefauthors}%
{Peterson}, J\BPBI R.%
, {Paerels}, F\BPBI B\BPBI S.%
, {Kaastra}, J\BPBI S.%
\ et al.\end{APACrefauthors}%
\unskip\
\newblock
\APACrefYearMonthDay{2001}{{\APACmonth{01}}}{},
\newblock
\unskip
\newblock
\APACjournalVolNumPages{A\&A}{365}{}{L104-L109}.
\newblock

\PrintBackRefs{\CurrentBib}

\bibitem [\protect \citeauthoryear {%
{Pinto}%
\ \protect \BOthers {.}}{%
{Pinto}%
\ \protect \BOthers {.}}{%
{\protect \APACyear {2018}}%
}]{%
Pinto2018b}
\APACinsertmetastar {%
Pinto2018b}%
\begin{APACrefauthors}%
{Pinto}, C.%
, {Bambic}, C\BPBI J.%
, {Sanders}, J\BPBI S.%
\ et al.\end{APACrefauthors}%
\unskip\
\newblock
\APACrefYearMonthDay{2018}{Nov}{},
\newblock
\unskip
\newblock
\APACjournalVolNumPages{\mnras}{480}{3}{4113-4123}.
\newblock

\PrintBackRefs{\CurrentBib}

\bibitem [\protect \citeauthoryear {%
{Pinto}%
\ \protect \BOthers {.}}{%
{Pinto}%
\ \protect \BOthers {.}}{%
{\protect \APACyear {2016}}%
}]{%
Pinto2016mnras}
\APACinsertmetastar {%
Pinto2016mnras}%
\begin{APACrefauthors}%
{Pinto}, C.%
, {Fabian}, A\BPBI C.%
, {Ogorzalek}, A.%
\ et al.\end{APACrefauthors}%
\unskip\
\newblock
\APACrefYearMonthDay{2016}{{\APACmonth{09}}}{},
\newblock
\unskip
\newblock
\APACjournalVolNumPages{MNRAS}{461}{}{2077-2084}.
\newblock

\PrintBackRefs{\CurrentBib}

\bibitem [\protect \citeauthoryear {%
{Pinto}%
\ \protect \BOthers {.}}{%
{Pinto}%
\ \protect \BOthers {.}}{%
{\protect \APACyear {2014}}%
}]{%
Pinto2014}
\APACinsertmetastar {%
Pinto2014}%
\begin{APACrefauthors}%
{Pinto}, C.%
, {Fabian}, A\BPBI C.%
, {Werner}, N.%
\ et al.\end{APACrefauthors}%
\unskip\
\newblock
\APACrefYearMonthDay{2014}{{\APACmonth{12}}}{},
\newblock
\unskip
\newblock
\APACjournalVolNumPages{A\&A}{572}{}{L8}.
\newblock

\PrintBackRefs{\CurrentBib}

\bibitem [\protect \citeauthoryear {%
{Pinto}%
\ \protect \BOthers {.}}{%
{Pinto}%
\ \protect \BOthers {.}}{%
{\protect \APACyear {2015}}%
}]{%
Pinto2015}
\APACinsertmetastar {%
Pinto2015}%
\begin{APACrefauthors}%
{Pinto}, C.%
, {Sanders}, J\BPBI S.%
, {Werner}, N.%
\ et al.\end{APACrefauthors}%
\unskip\
\newblock
\APACrefYearMonthDay{2015}{{\APACmonth{03}}}{},
\newblock
\unskip
\newblock
\APACjournalVolNumPages{A\&A}{575}{}{A38}.
\newblock

\PrintBackRefs{\CurrentBib}

\bibitem [\protect \citeauthoryear {%
{Roncarelli}%
\ \protect \BOthers {.}}{%
{Roncarelli}%
\ \protect \BOthers {.}}{%
{\protect \APACyear {2018}}%
}]{%
Roncarelli2018}
\APACinsertmetastar {%
Roncarelli2018}%
\begin{APACrefauthors}%
{Roncarelli}, M.%
, {Gaspari}, M.%
, {Ettori}, S.%
\ et al.\end{APACrefauthors}%
\unskip\
\newblock
\APACrefYearMonthDay{2018}{Oct}{},
\newblock
\unskip
\newblock
\APACjournalVolNumPages{\aap}{618}{}{A39}.
\newblock

\PrintBackRefs{\CurrentBib}

\bibitem [\protect \citeauthoryear {%
{Russell}%
, {McDonald}%
, {McNamara}%
, {Fabian}%
\BCBL {}\ \BBA {} {Nulsen}%
}{%
{Russell}%
\ \protect \BOthers {.}}{%
{\protect \APACyear {2017}}%
}]{%
Russell2017}
\APACinsertmetastar {%
Russell2017}%
\begin{APACrefauthors}%
{Russell}, H\BPBI R.%
, {McDonald}, M.%
, {McNamara}, B\BPBI R.%
, {Fabian}, A\BPBI C.%
\BCBL {}\ \BBA {} {Nulsen}, P\BPBI E\BPBI J\BPBI e\BPBI a.%
\end{APACrefauthors}%
\unskip\
\newblock
\APACrefYearMonthDay{2017}{{\APACmonth{02}}}{},
\newblock
\unskip
\newblock
\APACjournalVolNumPages{ApJ}{836}{}{130}.
\newblock

\PrintBackRefs{\CurrentBib}

\bibitem [\protect \citeauthoryear {%
{Sanders}%
\ \BBA {} {Fabian}%
}{%
{Sanders}%
\ \BBA {} {Fabian}%
}{%
{\protect \APACyear {2013}}%
}]{%
Sanders2013}
\APACinsertmetastar {%
Sanders2013}%
\begin{APACrefauthors}%
{Sanders}, J\BPBI S.%
\BCBT {}\ \BBA {} {Fabian}, A\BPBI C.%
\end{APACrefauthors}%
\unskip\
\newblock
\APACrefYearMonthDay{2013}{{\APACmonth{03}}}{},
\newblock
\unskip
\newblock
\APACjournalVolNumPages{MNRAS}{429}{}{2727-2738}.
\newblock

\PrintBackRefs{\CurrentBib}

\bibitem [\protect \citeauthoryear {%
{Sanders}%
\ \protect \BOthers {.}}{%
{Sanders}%
\ \protect \BOthers {.}}{%
{\protect \APACyear {2008}}%
}]{%
Sanders2008a}
\APACinsertmetastar {%
Sanders2008a}%
\begin{APACrefauthors}%
{Sanders}, J\BPBI S.%
, {Fabian}, A\BPBI C.%
, {Allen}, S\BPBI W.%
, {Morris}, R\BPBI G.%
, {Graham}, J.%
\BCBL {}\ \BBA {} {Johnstone}, R\BPBI M.%
\end{APACrefauthors}%
\unskip\
\newblock
\APACrefYearMonthDay{2008}{{\APACmonth{04}}}{},
\newblock
\unskip
\newblock
\APACjournalVolNumPages{MNRAS}{385}{}{1186-1200}.
\newblock

\PrintBackRefs{\CurrentBib}

\bibitem [\protect \citeauthoryear {%
{Sanders}%
, {Fabian}%
\BCBL {}\ \BBA {} {Smith}%
}{%
{Sanders}%
\ \protect \BOthers {.}}{%
{\protect \APACyear {2011}}%
}]{%
Sanders2011b}
\APACinsertmetastar {%
Sanders2011b}%
\begin{APACrefauthors}%
{Sanders}, J\BPBI S.%
, {Fabian}, A\BPBI C.%
\BCBL {}\ \BBA {} {Smith}, R\BPBI K.%
\end{APACrefauthors}%
\unskip\
\newblock
\APACrefYearMonthDay{2011}{{\APACmonth{01}}}{},
\newblock
\unskip
\newblock
\APACjournalVolNumPages{MNRAS}{410}{}{1797-1812}.
\newblock

\PrintBackRefs{\CurrentBib}

\bibitem [\protect \citeauthoryear {%
{Sanders}%
, {Fabian}%
, {Smith}%
\BCBL {}\ \BBA {} {Peterson}%
}{%
{Sanders}%
\ \protect \BOthers {.}}{%
{\protect \APACyear {2010}}%
}]{%
Sanders2010}
\APACinsertmetastar {%
Sanders2010}%
\begin{APACrefauthors}%
{Sanders}, J\BPBI S.%
, {Fabian}, A\BPBI C.%
, {Smith}, R\BPBI K.%
\BCBL {}\ \BBA {} {Peterson}, J\BPBI R.%
\end{APACrefauthors}%
\unskip\
\newblock
\APACrefYearMonthDay{2010}{{\APACmonth{02}}}{},
\newblock
\unskip
\newblock
\APACjournalVolNumPages{MNRAS}{402}{}{L11-L15}.
\newblock

\PrintBackRefs{\CurrentBib}

\bibitem [\protect \citeauthoryear {%
{Simionescu}%
\ \protect \BOthers {.}}{%
{Simionescu}%
\ \protect \BOthers {.}}{%
{\protect \APACyear {2019}}%
}]{%
Simionescu2019a}
\APACinsertmetastar {%
Simionescu2019a}%
\begin{APACrefauthors}%
{Simionescu}, A.%
, {Nakashima}, S.%
, {Yamaguchi}, H.%
\ et al.\end{APACrefauthors}%
\unskip\
\newblock
\APACrefYearMonthDay{2019}{Feb}{},
\newblock
\unskip
\newblock
\APACjournalVolNumPages{\mnras}{483}{2}{1701-1721}.
\newblock

\PrintBackRefs{\CurrentBib}

\bibitem [\protect \citeauthoryear {%
{Simionescu}%
\ \protect \BOthers {.}}{%
{Simionescu}%
\ \protect \BOthers {.}}{%
{\protect \APACyear {2009}}%
}]{%
Simionescu2009a}
\APACinsertmetastar {%
Simionescu2009a}%
\begin{APACrefauthors}%
{Simionescu}, A.%
, {Werner}, N.%
, {B{\"o}hringer}, H.%
\ et al.\end{APACrefauthors}%
\unskip\
\newblock
\APACrefYearMonthDay{2009}{Jan}{},
\newblock
\unskip
\newblock
\APACjournalVolNumPages{\aap}{493}{2}{409-424}.
\newblock

\PrintBackRefs{\CurrentBib}

\bibitem [\protect \citeauthoryear {%
{Tamura}%
\ \protect \BOthers {.}}{%
{Tamura}%
\ \protect \BOthers {.}}{%
{\protect \APACyear {2001}}%
}]{%
Tamura2001b}
\APACinsertmetastar {%
Tamura2001b}%
\begin{APACrefauthors}%
{Tamura}, T.%
, {Bleeker}, J\BPBI A\BPBI M.%
, {Kaastra}, J\BPBI S.%
\ et al.\end{APACrefauthors}%
\unskip\
\newblock
\APACrefYearMonthDay{2001}{Nov}{},
\newblock
\unskip
\newblock
\APACjournalVolNumPages{\aap}{379}{}{107-114}.
\newblock

\PrintBackRefs{\CurrentBib}

\bibitem [\protect \citeauthoryear {%
{Tamura}%
\ \protect \BOthers {.}}{%
{Tamura}%
\ \protect \BOthers {.}}{%
{\protect \APACyear {2003}}%
}]{%
Tamura2003}
\APACinsertmetastar {%
Tamura2003}%
\begin{APACrefauthors}%
{Tamura}, T.%
, {Kaastra}, J\BPBI S.%
, {Makishima}, K.%
\ et al.\end{APACrefauthors}%
\unskip\
\newblock
\APACrefYearMonthDay{2003}{Feb}{},
\newblock
\unskip
\newblock
\APACjournalVolNumPages{\aap}{399}{}{497-504}.
\newblock

\PrintBackRefs{\CurrentBib}

\bibitem [\protect \citeauthoryear {%
{Urban}%
\ \protect \BOthers {.}}{%
{Urban}%
\ \protect \BOthers {.}}{%
{\protect \APACyear {2017}}%
}]{%
Urban2017}
\APACinsertmetastar {%
Urban2017}%
\begin{APACrefauthors}%
{Urban}, O.%
, {Werner}, N.%
, {Allen}, S\BPBI W.%
\ et al.\end{APACrefauthors}%
\unskip\
\newblock
\APACrefYearMonthDay{2017}{Oct}{},
\newblock
\unskip
\newblock
\APACjournalVolNumPages{\mnras}{470}{4}{4583-4599}.
\newblock

\PrintBackRefs{\CurrentBib}

\bibitem [\protect \citeauthoryear {%
{Werner}%
, {B{\"o}hringer}%
\BCBL {}\ \protect \BOthers {.}}{%
{Werner}%
, {B{\"o}hringer}%
\BCBL {}\ \protect \BOthers {.}}{%
{\protect \APACyear {2006}}%
}]{%
Werner2006b}
\APACinsertmetastar {%
Werner2006b}%
\begin{APACrefauthors}%
{Werner}, N.%
, {B{\"o}hringer}, H.%
, {Kaastra}, J\BPBI S.%
, {de Plaa}, J.%
, {Simionescu}, A.%
, {Vink}, J.%
\BCBL {}\ \BOthersPeriod {.}\end{APACrefauthors}%
\unskip\
\newblock
\APACrefYearMonthDay{2006}{Nov}{},
\newblock
\unskip
\newblock
\APACjournalVolNumPages{\aap}{459}{2}{353-360}.
\newblock

\PrintBackRefs{\CurrentBib}

\bibitem [\protect \citeauthoryear {%
{Werner}%
, {de Plaa}%
\BCBL {}\ \protect \BOthers {.}}{%
{Werner}%
, {de Plaa}%
\BCBL {}\ \protect \BOthers {.}}{%
{\protect \APACyear {2006}}%
}]{%
Werner2006a}
\APACinsertmetastar {%
Werner2006a}%
\begin{APACrefauthors}%
{Werner}, N.%
, {de Plaa}, J.%
, {Kaastra}, J\BPBI S.%
\ et al.\end{APACrefauthors}%
\unskip\
\newblock
\APACrefYearMonthDay{2006}{{\APACmonth{04}}}{},
\newblock
\unskip
\newblock
\APACjournalVolNumPages{A\&A}{449}{}{475-491}.
\newblock

\PrintBackRefs{\CurrentBib}

\bibitem [\protect \citeauthoryear {%
{Werner}%
\ \protect \BOthers {.}}{%
{Werner}%
\ \protect \BOthers {.}}{%
{\protect \APACyear {2009}}%
}]{%
Werner2009}
\APACinsertmetastar {%
Werner2009}%
\begin{APACrefauthors}%
{Werner}, N.%
, {Zhuravleva}, I.%
, {Churazov}, E.%
\ et al.\end{APACrefauthors}%
\unskip\
\newblock
\APACrefYearMonthDay{2009}{{\APACmonth{09}}}{},
\newblock
\unskip
\newblock
\APACjournalVolNumPages{MNRAS}{398}{}{23-32}.
\newblock

\PrintBackRefs{\CurrentBib}

\bibitem [\protect \citeauthoryear {%
{Zhuravleva}%
\ \protect \BOthers {.}}{%
{Zhuravleva}%
\ \protect \BOthers {.}}{%
{\protect \APACyear {2014}}%
}]{%
Zhuravleva2014}
\APACinsertmetastar {%
Zhuravleva2014}%
\begin{APACrefauthors}%
{Zhuravleva}, I.%
, {Churazov}, E.%
, {Schekochihin}, A\BPBI A.%
, {Allen}, S\BPBI W.%
, {Ar{\'e}valo}, P.%
\BCBL {}\ \BOthersPeriod {.}\end{APACrefauthors}%
\unskip\
\newblock
\APACrefYearMonthDay{2014}{{\APACmonth{11}}}{},
\newblock
\unskip
\newblock
\APACjournalVolNumPages{Nature}{515}{}{85-87}.
\newblock

\PrintBackRefs{\CurrentBib}

\end{thebibliography}

\end{document}